\documentclass[12pt,10pt,aps,prl,longbibliography,twocolumn]{revtex4-1}
\usepackage[latin9]{inputenc}
\usepackage{amsmath}
\usepackage{amssymb}
\usepackage{graphicx}

\makeatletter

\@ifundefined{textcolor}{}
{%
\definecolor{BLACK}{gray}{0}
\definecolor{WHITE}{gray}{1}
\definecolor{RED}{rgb}{1,0,0}
\definecolor{GREEN}{rgb}{0,1,0}
\definecolor{BLUE}{rgb}{0,0,1}
\definecolor{CYAN}{cmyk}{1,0,0,0}
\definecolor{MAGENTA}{cmyk}{0,1,0,0}
\definecolor{YELLOW}{cmyk}{0,0,1,0}
}

\makeatother

\begin{document}

\title{Fate of the false vacuum: towards realization with ultra-cold atoms}

\author{O. Fialko$^{1}$, B. Opanchuk$^{2}$, A. I. Sidorov$^{2}$, P. D.
Drummond$^{2}$, J. Brand$^{3}$ }

\affiliation{$^{1}$Institute of Natural and Mathematical Sciences and Centre
for Theoretical Chemistry and Physics, Massey University, Auckland,
New Zealand}

\affiliation{$^{2}$Centre for Quantum and Optical Science, Swinburne University
of Technology, Melbourne 3122, Australia}

\affiliation{$^{3}$Dodd-Walls Centre for Photonic and Quantum Technologies, New
Zealand Institute for Advanced Study, and Centre for Theoretical Chemistry
and Physics, Massey University, Auckland, New Zealand}
\begin{abstract}
Quantum decay of a relativistic scalar field from a false vacuum is
a fundamental idea in quantum field theory. It is relevant to models
of the early Universe, where the nucleation of bubbles gives rise
to an inflationary universe and the creation of matter. Here we propose
a laboratory test using an experimental model of an ultra-cold spinor
Bose gas. A false vacuum for the relative phase of two spin components,
serving as the unstable scalar field, is generated by means of a modulated
radio-frequency coupling of the spin components. Numerical simulations
demonstrate the spontaneous formation of true vacuum bubbles with
realistic parameters and time-scales.
\end{abstract}


\maketitle
As proposed by Coleman in a seminal paper~\cite{Coleman1977}, the
false vacuum is a metastable state of the relativistic scalar field
that can decay by quantum tunneling, locally forming bubbles of true
vacuum that expand at the speed of light. It has a close analogy with
the ubiquitous phenomenon of bubble nucleation during a first order
phase transition in condensed matter~\cite{Langer1967a}, e.g.\ the
spontaneous creation of vapor bubbles in superheated water~\cite{Schmelzer2005}.
Applied to a quantum field such as the inflaton or Higgs field, bubble
nucleation is an event of cosmological significance in some early
universe models. Indeed, the Coleman decay scenario of the inflaton
field features prominently in the theory of eternal inflation~\cite{Vilenkin1983,Guth2007},
where bubbles continuously nucleating from a false vacuum grow into
separate universes, each subsequently undergoing exponential growth
of space~\cite{Coleman1980}. This scenario, which could potentially
explain the value of the cosmological constant by the anthropic principle,
is currently being tested against observational evidence in astrophysical
experiments~\cite{Feeney2011-inflation,Bousso2013}. For an observer
inside the bubble, the tunneling event~\textemdash{} occurring in
the observer's past~\textemdash{} appears like a cosmological ``big-bang'',
prior to inflation.

From a theoretical point of view, quantum tunneling from a false vacuum
is a problem that can only be solved approximately~\cite{Coleman1977,Callan1977}
(except for simplified models~\cite{Drummond1989}) due to the exponential
complexity of quantum field dynamics. This motivates the search for
an analog quantum system that is accessible to experimental scrutiny,
to test these models. The utility of such experiments, which complement
astrophysical investigations, is that they would provide data that
allow verification of widely used approximations inherent in current
theories~\cite{Georgescu2014}.

Here we demonstrate how to use an ultra-cold atomic two-component
Bose-Einstein condensate (BEC) as a quantum simulator that generates
a decaying, relativistic false vacuum. Quantum field dynamics occurs
for the relative phase of two spin components that are linearly coupled
by a radio-frequency field. In this proposal the speed of sound in
the condensate models the speed of light, and the ``universe'' is
less than a millimeter across. Domains of true vacuum are observable
using interferometric techniques~\cite{Egorov2011} over millisecond
time-scales with realistic parameters.

Modulating the radio-frequency coupling in time allows one to create
a metastable vacuum from an otherwise unstable one~\cite{Opanchuk2013-early-universe}
following Kapitza's famous idea for stabilizing the unstable point
of a pendulum by rocking the pivot point~\cite{Kapitza1951-JETP}.
Our proposal requires repulsive intra-component interactions to dominate
over inter-component interactions. To achieve this, we have identified
a Feshbach resonance of $^{41}$K with a zero crossing for the inter-component
$s$-wave scattering. Similar couplings in lower dimensions can also
be achieved with a transverse double-well potential~\cite{Kaurov2005,Gritsev2007,Brand2009}. 

Previous work on ultra-cold atom analog models of the early universe
has focused on the expansion of space-time~\cite{Fischer2004,Menicucci2010}
and the formation of oscillons~\cite{Neuenhahn2012-phase-structures,Amin2012}.
While interesting cosmological analogs have been explored in liquid
He~\cite{Bauerle1996,Volovik2009}, its use as a quantum simulator
is hampered by the limited tuneablility of physical parameters and
the phenomenological nature of available theoretical models. False
vacuum decay models have been successfully applied to the quantum
nucleation of phase transitions of liquid He~\cite{Satoh1992,Tye2011}
in a non-relativisitc context, but the nucleation and decay from a
relativistic false vacuum has not yet been realized in a laboratory
experiment. In this Letter we propose an implementation of Coleman's
model of quantum decay from a relativistic false vacuum with tuneable
microscopic parameters. We simulate the quantum dynamics of the coupled
Bose fields in the truncated Wigner approximation (TWA)~\cite{Steel1998,Sinatra2002}
and demonstrate how the resulting evolution can be imaged in one,
two or three space dimensions using optical trapping. This shows the
feasibility of a table-top experiment, and illustrates how the expected
bubble nucleation dynamics depends on the dimensionality of space.

The dynamics of the scalar field $\phi$ in Coleman's model~\cite{Coleman1977}
is given by the equation 
\begin{equation}
\partial_{t}^{2}\phi-c^{2}\nabla^{2}\phi=-\partial_{\phi}V(\phi),\label{eq:Hubble-1}
\end{equation}
where $c$ is the speed of light, and the potential $V(\phi)$ has
a metastable local minimum separated from a true vacuum by a barrier.
We emulate this equation with a pseudo spin-$1/2$ BEC, where the
speed of light is replaced by the speed of sound in the BEC, the relative
phase between two spin components assumes the role of the scalar field
$\phi$, and the shape of the potential $V(\phi)$ is tunable. In
addition there is an adjustable coupling to phonon degrees of freedom
in our system, which serves to damp the dynamics. Our numerical simulations
confirm the expected features of quantum tunneling dynamics with dissipation~\cite{Caldeira1981}.
By addressing a radio-frequency transition between the spin components,
the false vacuum initial state can be prepared and the final state
read out by interferometry.

We consider a two-component BEC of atoms with mass $m$ and a linear
coupling $\nu$ realized by a radio-frequency field. Atoms with the
same spin interact via a point-like potential with strength $g$.
The Hamiltonian reads 

\begin{eqnarray}
\hat{H} & = & \int d{\bf r}\hat{\psi}_{\sigma}^{\dagger}({\bf r})\left[-\frac{\hbar^{2}\nabla^{2}}{2m}-\mu\right]\hat{\psi}_{\sigma}({\bf r})-\nu\int d{\bf r}\hat{\psi}_{\sigma}^{\dagger}({\bf r})\hat{\psi}_{\bar{\sigma}}({\bf r})\nonumber \\
 & + & \frac{g}{2}\int d{\bf r}\hat{\psi}_{\sigma}^{\dagger}({\bf r})\hat{\psi}_{\sigma}^{\dagger}({\bf r})\hat{\psi}_{\sigma}({\bf r})\hat{\psi}_{\sigma}({\bf r}),\label{eq:Hamiltonian}
\end{eqnarray}
where summation over spin indices $\sigma\in\{-,+\}$ is implied.
The Bose fields satisfy the usual commutation relations $\left[\hat{\psi}_{\sigma}({\bf r}),\hat{\psi}_{\bar{\sigma}}^{\dagger}({\bf r}')\right]=i\delta_{\sigma\bar{\sigma}}\delta({\bf r}-{\bf r'})$.
We introduce the quantum partition function ${\cal Z}=\int{\cal D}(\psi^{\ast},\psi){\rm e^{-S[\psi^{\ast},\psi]}}$~\cite{Atland2010},
where $S[\psi^{\ast},\psi]=\int d{\bf s}\left[\psi_{\sigma}^{\ast}\partial_{\tau}\psi_{\sigma}+H(\psi^{\ast},\psi)\right]$.
Here, ${\bf s}=(\tau,{\bf r})$ is a $d+1$ vector, $\tau=it/\hbar\in[0,\beta]$
is imaginary time, $\psi_{\sigma}(\tau,{\bf r})$ is a complex field
subject to the periodic boundary condition $\psi_{\sigma}({\rm {\bf \beta,r}})=\psi_{\sigma}({\rm 0,{\bf r}})$.
We look first for a static solution to identify vacua. This amounts
to replacing $\psi_{\sigma}=\psi_{0}={\rm const}$ in the saddle-point
approximation $\delta S/\delta\psi_{\sigma}=0$. For $\nu>0$ we obtain
the stable $|\psi_{0}|^{2}=(\mu+\nu)/g$ and unstable $|\psi_{0}|^{2}=(\mu-\nu)/g$
vacua with the two Bose gases being in phase and out-of-phase respectively.
Let us introduce new field variables by $\psi_{\sigma}({\bf s})=\rho_{\sigma}^{1/2}({\bf s})e^{i\phi_{\sigma}({\bf s})}$,
where $\rho_{\sigma}({\bf s})=\rho_{0}+\delta\rho_{\sigma}({\bf s})$
and $\rho_{0}=|\psi_{0}|^{2}$. The variables $\delta\rho_{\sigma}$
and $\phi_{\sigma}$ parametrize the deviation of the Bose fields
from a vacuum. Substituting this parametrization into the action,
we obtain

\begin{eqnarray}
S(\rho,\phi) & \approx & \int d{\bf s}\left[i\delta\rho_{\sigma}\partial_{\tau}\phi_{\sigma}+\frac{\hbar^{2}\rho_{0}}{2m}(\nabla\phi_{\sigma})^{2}\right.\nonumber \\
 & + & \frac{(2g\rho_{0}+\nu)\delta\rho_{\sigma}^{2}}{4\rho_{0}}+\frac{\hbar^{2}(\nabla\delta\rho_{\sigma})^{2}}{8m\rho_{0}}-\frac{\nu}{2\rho_{0}}\delta\rho_{1}\delta\rho_{2}\nonumber \\
 & - & \left.2\nu\rho_{0}\cos(\phi_{a})-\nu\cos(\phi_{a})\delta\rho_{\sigma}\right].\label{eq:S_rho_fi}
\end{eqnarray}
Here $\phi_{a}=\phi_{+}-\phi_{-}$ is the relative phase. Through
elimination of slowly varying density fluctuations following the standard
technique~\cite{Atland2010}, an effective field theory for the phase
difference relevant for energies below $\hbar\xi c$ $(\xi=\hbar/\sqrt{2mg\rho_{0}}$
is the BEC healing length) is found 

\begin{eqnarray}
S(\phi_{a}) & = & \frac{1}{4g}\int d{\bf s}[(\partial_{\tau}\phi_{a})^{2}+\hbar^{2}c^{2}(\nabla\phi_{a})^{2}+2\hbar^{2}V(\phi_{a})]\nonumber \\
 & + & \frac{4\nu^{2}}{g}\int d{\bf s}\int d{\bf s}^{\prime}\cos\phi_{a}({\bf s})\cos\phi_{a}({\bf s}^{\prime}){\cal G}({\bf s}-{\bf s}^{\prime}),
\end{eqnarray}
where the field potential $V(\phi_{a})=-4\nu g\rho_{0}\cos\phi_{a}(1+\nu\cos\phi_{a}/4g\rho_{0})/\hbar^{2}$
and $c=\sqrt{g\rho_{0}/m}$ is the speed of sound. The non-local kernel
${\cal G}(\tau,{\bf r})=(\beta V)^{-1}\sum_{\omega_{n}>0}\sum_{{\bf k}>0}e^{-i(\omega_{n}\tau+{\bf k{\bf r}})}\omega_{n}^{2}/[\omega_{n}^{2}+(c\hbar)^{2}{\bf k}{}^{2}]$
is expressed through summation over Fourier momentum ${\bf k}$ and
Matsubara frequencies $\omega_{n}=2\pi n/\beta$. The new action is
similar to the one studied in Ref.~\cite{Caldeira1981}, where the
effect of dissipation on quantum dynamics was explored. Following
Ref.~\cite{Caldeira1981} we obtain the following equation of motion
for the relative phase: 
\begin{equation}
\partial_{t}^{2}\phi_{a}-c^{2}\nabla^{2}\phi_{a}+\frac{4\nu^{2}\xi}{\hbar^{2}c}\partial_{t}\phi_{a}=-\partial_{\phi_{a}}V(\phi_{a}).\label{eq:SG_fulla}
\end{equation}
It is similar to Eq.~\eqref{eq:Hubble-1} with a new friction-like
third term, due to coupling with density fluctuations. We note that
in cosmological models such friction-like behavior may occur due to
a homogeneous expansion of space, as commonly described by the Hubble
constant~\cite{Liddle2000}.

\begin{figure}
\includegraphics[width=0.9\columnwidth]{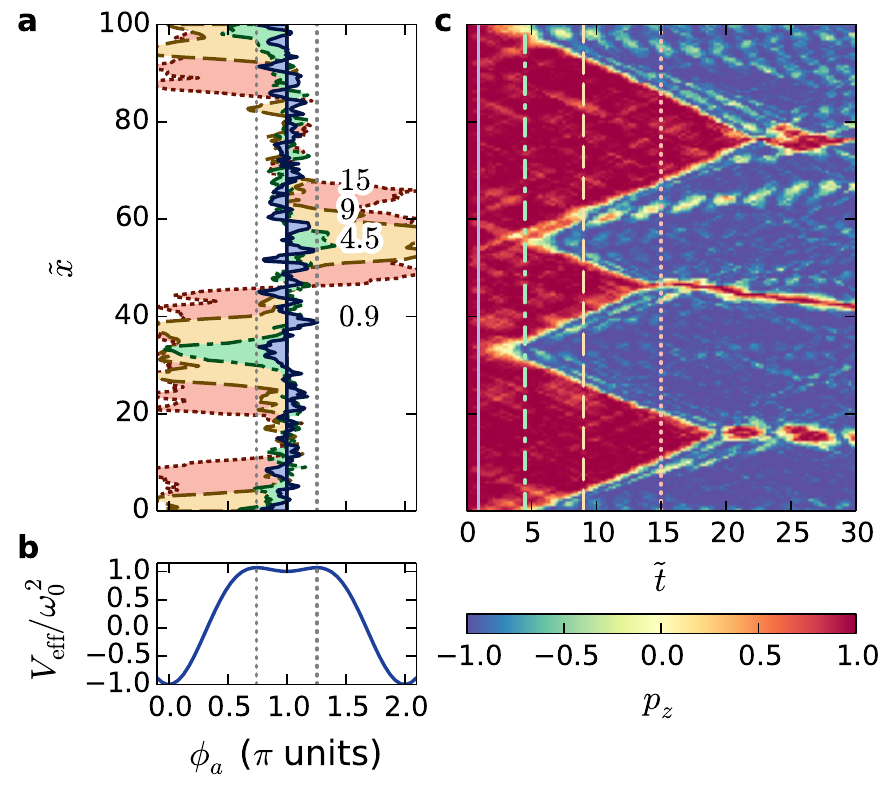}

\protect\caption{\textbf{\label{fig:1d-bubbles}}Decay of the false vacuum in 1D. A
single-trajectory simulation of the false vacuum decay in 1D with
$N_{\mathrm{grid}}=256$ and dimensionless parameters $\lambda=1.2$,
$\tilde{\omega}=50$, $\tilde{\nu}=0.01$, $\tilde{\rho}_{0}=200$,
$a_{11}=59.5$, $a_{22}=60.5$, $a_{12}=0$ (corresponding to a two-component
$^{41}$K condensate in a ring trap with $N=4\times10^{4}$, trap
circumference $L=254\,\mathrm{\mu m}$, transverse frequency $\omega_{\perp}=2\pi\times1913\,\mathrm{Hz}$,
observation time $T=24.9\,\mathrm{ms}$, oscillator amplitude $\Omega=2\pi\times9.56\,\mathrm{Hz}$,
frequency $\omega=2\pi\times9.56\,\mathrm{kHz}$ and modulation $\delta=0.085$).
\textbf{a}, Example of bubble formation: the spinor Bose field is
initially in a false vacuum ($\tilde{t}=0.9$, blue solid); quantum
fluctuations cause the field to tunnel out ($\tilde{t}=4.5$, green
dash-dotted); three bubbles are formed in true vacua ($\tilde{t}=9$,
yellow dashed); they grow until one bubble meets another bubble in
the second minimum, creating a domain wall ($\tilde{t}=15$, pink
dotted). \textbf{b}, Effective field potential (dotted lines mark
the potential maxima). \textbf{c}, Time evolution of the relative
number density difference $p_{z}$ after a $\pi/2$ rotation, which
converts the relative phase into a population difference.}
\end{figure}

By varying the tunnel coupling $\nu$ periodically in time, it is
possible to alter the effective potential so that $\phi_{a}=\pi$
becomes a local minimum, which corresponds to a false vacuum in the
sense of Coleman. We consider rapid oscillations of the tunnel coupling
$\nu_{t}=\nu+\delta\hbar\omega\cos(\omega t)$, where the frequency
of oscillations is $\omega\gg\omega_{0}\equiv2\sqrt{\nu g\rho_{0}}/\hbar$.
Following Kapitza~\cite{Kapitza1951-JETP} the field $\phi_{a}$
may be viewed now as a superposition of a slow component $\phi_{0}$
and rapid oscillations. Averaging rapid oscillations in time yields
an equation of the form~\eqref{eq:SG_fulla} with $\phi_{a}\rightarrow\phi_{0}$
and $V(\phi_{a})\rightarrow V_{{\rm eff}}(\phi_{0})=-\omega_{0}^{2}\left[\cos\phi_{0}-0.5\lambda^{2}\sin^{2}\phi_{0}\right]$.
This potential, shown in Fig.~\ref{fig:1d-bubbles}b, develops a
local minimum at $\phi_{0}=\pi$ if $\lambda=\delta\hbar\omega_{0}/\sqrt{2}\nu>1$,
which corresponds to a false vacuum. Multiple equivalent true vacua
occur at the global minima with $\phi_{0}=0,2\pi,4\pi,\ldots$ 

We perform stochastic numerical simulations on the full BEC model~\eqref{eq:Hubble-1}
to investigate the bubble nucleation numerically and compare the results
with the predictions of the effective theory developed above. The
TWA, where a quantum state is represented by a stochastic phase space
distribution of trajectories following the Gross-Pitaevskii equation~\cite{Steel1998,Sinatra2002},
enables one to simulate the entire experimental model of a three-dimensional
coupled spinor BEC, in the limit of large occupation numbers per mode.
This method is already known to accurately simulate BEC interferometric
experiments down to the quantum noise level~\cite{Egorov2011}. 

Our initial state construction then proceeds by assuming each mode
is initially in a coherent state. The corresponding Wigner distribution
is a Gaussian in phase space. The primary physical effect of the noise
is to allow spontaneous tunneling and scattering processes that are
disallowed in pure Gross-Pitaevskii theory. Quantum noise is added
to the classical false vacuum state as $\psi({\bf r})=\psi_{0}({\bf r})+\sum_{j=1}^{M}\alpha_{j}\exp(i{\bf k}_{j}{\bf r})/\sqrt{V}$.
Here $\alpha_{j}$ are complex Gaussian variables with $\overline{\alpha_{j}^{\ast}\alpha_{i}}=\delta_{ij}/2$,
thus sampling fluctuations of the false vacuum\@. Quantum tunneling
for a shallow potential well is equivalent to an activation process
caused by the vacuum fluctuations of the quantum field, represented
by the initial fluctuations of the Wigner phase-space representation.

The TWA is a truncation of the expansion in the powers of $M/N$ up
to and including the terms of order $1$~\cite{Sinatra2002}. Therefore,
the number of modes $M$ is chosen to represent the physical system,
while being much smaller than the number of atoms $N$. For the 1D
and the 3D simulations, $M/N\equiv N_{\mathrm{grid}}^{d}/\left(\tilde{L}^{d}\tilde{\rho}_{0}\right)\approx10^{-2}$,
and for the 2D simulation $M/N\approx3\times10^{-2}$, where $N_{\mathrm{grid}}$
is the number of grid points in one dimension. 

We propagate this state in real time by solving the time-dependent
coupled equations 
\begin{equation}
i\hbar\partial_{t}\psi_{j}=\left[-\frac{\hbar^{2}}{2m}\nabla^{2}-\mu+g\left(|\psi_{j}|^{2}-\frac{M}{V}\right)\right]\psi_{j}-\nu_{t}\psi_{3-j},\label{eq:GP}
\end{equation}
where index $j=1,2$, and with the coupling $\nu_{t}$ modulated in
time. 

The results of the simulations are shown in Fig.~\ref{fig:1d-bubbles}.
The single trajectory dynamics shown here features the creation of
three bubbles. Collisions of bubbles result either in the creation
of localized long-lived oscillating structures known as oscillons~\cite{Vachaspati2006,Amin2012,Su2014a},
or domain walls if the colliding bubbles belong to topologically distinct
vacua.

\begin{figure}
\includegraphics[width=0.9\columnwidth]{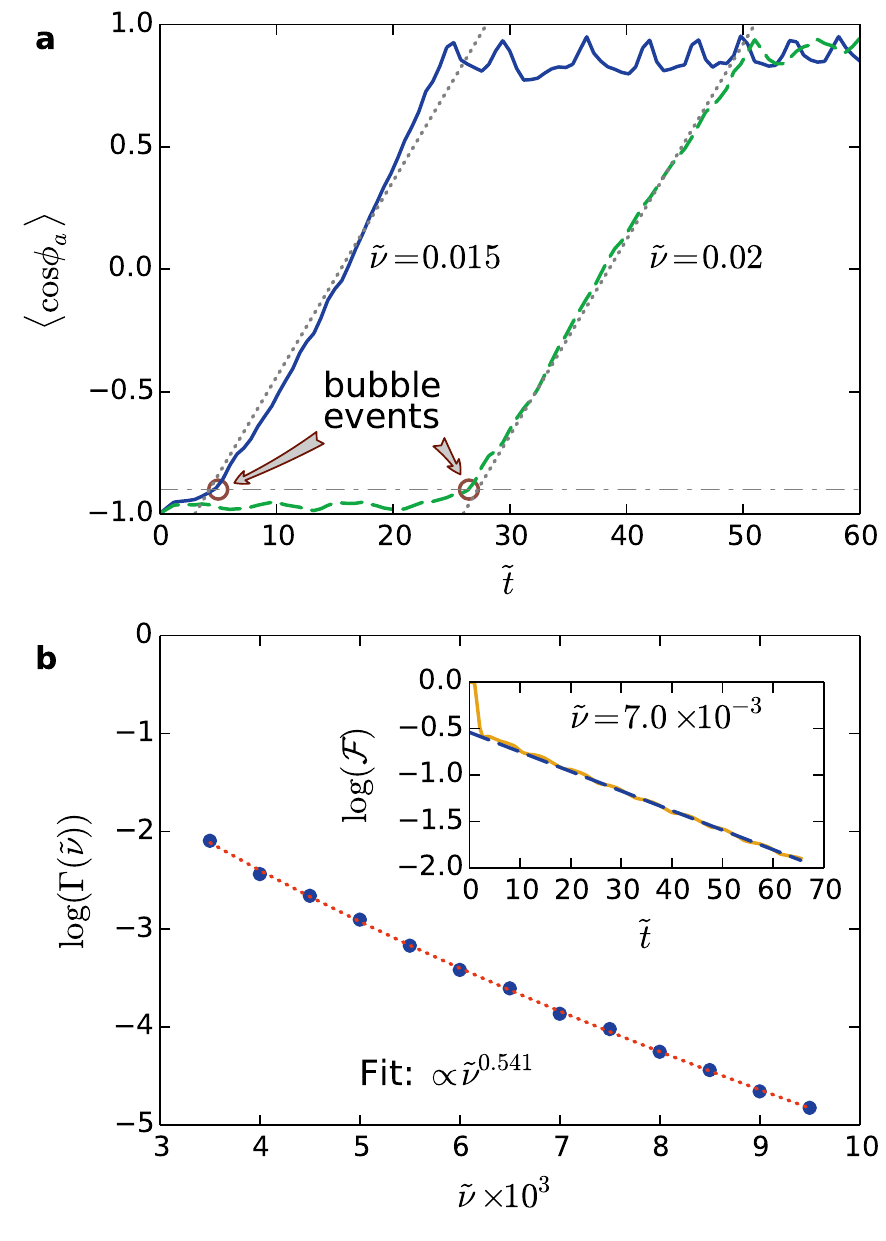}\protect\caption{\label{fig:1d-fitting}Bubble nucleation probability. \textbf{a,}
Single trajectories of the relative phase at different $\tilde{\nu}$,
where $\langle\cos\phi_{a}\rangle=\frac{1}{L}\int_{0}^{L}dx\cos\phi_{a}(x)$.
Initially the field is trapped in a false vacuum. A bubble appears
via quantum tunneling at $t\omega_{0}=5$ (blue line), $t\omega_{0}=26$
(green line). The tunneling time is longer for larger couplings $\nu$,
consistent with dissipative slowing down of tunneling~\cite{Caldeira1981}.
The bubble grows at the speed of sound ($1$ in our units). \textbf{b}
(inset), Probability of bubble nucleation for $\tilde{\nu}=8\times10^{-3}$
and its exponential fit. \textbf{b,} Dependence of the tunneling rate
$\Gamma$ on the coupling $\tilde{\nu}$ for $\lambda=1.3$. This
is extracted from the survival probability of the bubble nucleation,
which behaves as ${\cal F}=\exp(-\Gamma t)$. For this relatively
shallow effective potential, we extract that $\Delta B\propto\tilde{\nu}^{0.541}$.}
\end{figure}

To quantify the tunneling process, we calculate the probability that
the system has not yet decayed at time $t$. At long time scales it
should behave as ${\cal F}=\exp(-\Gamma t)$~\cite{Takagi2006},
where $\Gamma$ is the decay rate from the false vacuum. From the
probability of bubble creation over time ${\cal P}(t)$, the ``survival''
probability can be calculated as ${\cal F}=1-\int_{0}^{t}{\cal P}(t')dt'$
and the decay rate can be extracted. In the weak tunneling limit it
can be written in the form $\Gamma=A\exp(-B/\hbar)$ and the coefficients
$A$ and $B$ were calculated in Refs.~\cite{Coleman1977,Callan1977}
in limiting cases. Our numerical simulations are far from that to
allow an experiment on reasonable time-scale. In the following we
focus on the damping term, the third term in Eq.~\eqref{eq:SG_fulla},
which suppresses tunneling leading to a correction to $B$, $\Delta B\propto\tilde{\nu}^{3/2}$~\cite{Caldeira1981}
in the weak tunneling regime. 

\begin{figure}
\includegraphics[width=0.9\columnwidth]{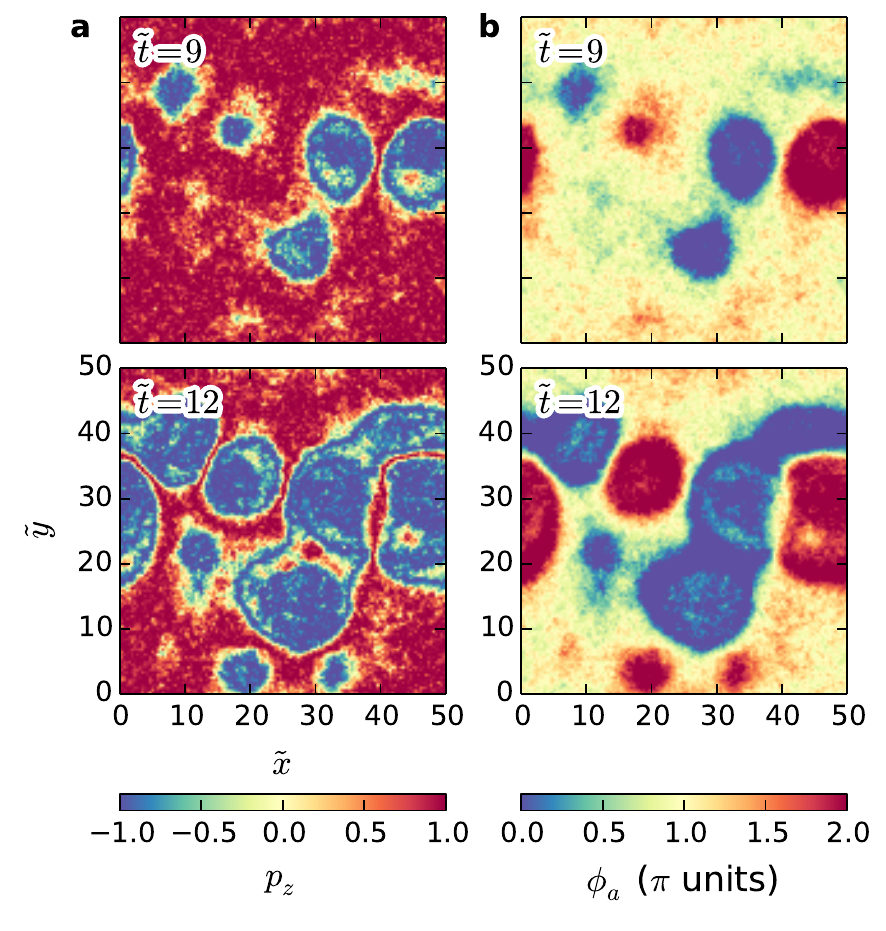}\protect\caption{\label{fig:2d-bubbles}Bubble formation in 2D. Snapshots representing
the number density difference (\textbf{a}) and the phase difference
(\textbf{b}) with $N_{\mathrm{grid}}=128$ and dimensionless parameters
$\lambda=1.1$, $\tilde{\omega}=50$, $\tilde{\nu}=0.005$, $\tilde{\rho}_{0}=200$,
$a_{11}=59.5$, $a_{22}=60.5$, $a_{12}=0$. Bubbles are seeded through
quantum tunnelling and grow ($\tilde{t}=9$) forming domain walls
between bubbles with two distinct phases of 0 and $2\pi$ ($\tilde{t}=12$).}
\end{figure}

Our TWA approach is expected to yield accurate predictions for the
relatively shallow effective potentials necessary for tunneling over
laboratory time-scales~\cite{Drummond1989}. In Fig.~\ref{fig:1d-fitting}
we present the scaling of the tunneling rate of bubbles for $\lambda=1.3$.
We find that the corresponding $\Delta B\propto\tilde{\nu}^{0.5}$.
We checked numerically that by increasing $\lambda$ the exponent
also increases. However, our approach is not valid for larger values
of $\lambda$, and this is also less accessible experimentally. The
observed behavior provides strong evidence of a quantum tunneling
process. 
\begin{figure}[t]
\includegraphics[width=1\columnwidth]{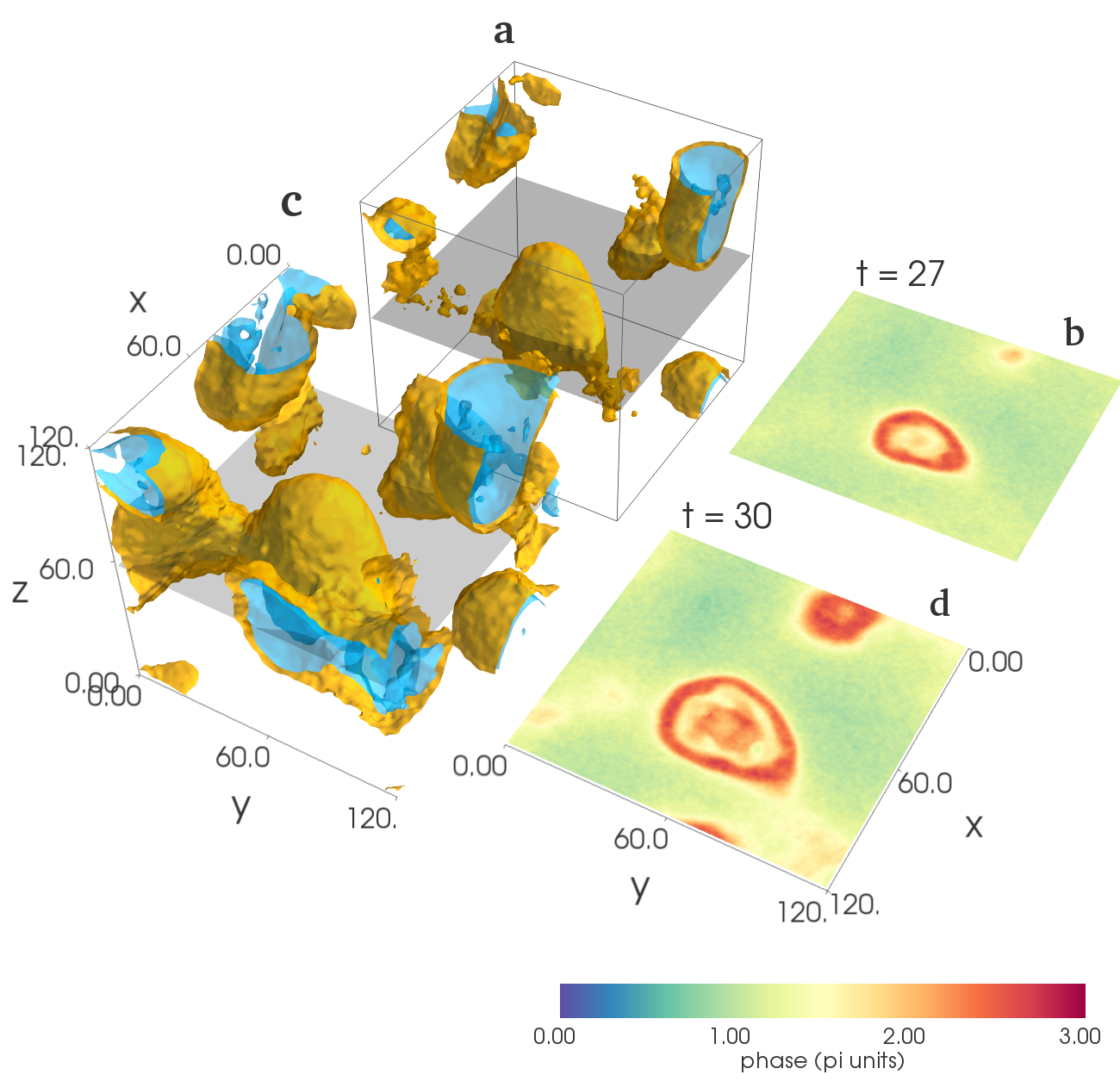}

\protect\caption{\label{fig:3d-bubbles}Bubble formation in 3D. Simulation of a single
three-dimensional random trajectory with $N_{\mathrm{grid}}=128$
and dimensionless parameters $\lambda=1.01$, $\tilde{\omega}=50$,
$\tilde{\nu}=0.05$, $\tilde{\rho}_{0}=100$, $a_{11}=59.5$, $a_{22}=60.5$,
$a_{12}=0$. This corresponds to a two-component $^{41}$K condensate
with realistic parameters in a uniform box trap~\cite{Gaunt2013}
with $N=1.72\times10^{8}$, trap size $L=95.8\,\mathrm{\mu m}$, observation
time $T=7.4\,\mathrm{ms}$, oscillator amplitude $\Omega=2\pi\times96.9\,\mathrm{Hz}$,
frequency $\omega=2\pi\times43.3\,\mathrm{kHz}$ and modulation $\delta=0.16$.
\textbf{a, c,} The 3D outline of relative phases of 0.2 (yellow) and
0.8 (blue) converted into a relative number density difference $p_{z}$.
\textbf{b, d, }The phase difference in a 2D slice to show the internal
phase variation near the bubble walls (location is marked with gray
on the corresponding 3D plots). Dimensionless time is $\tilde{t}=27$
for \textbf{a, b}, and $\tilde{t}=30$ for \textbf{c, d}.}
\end{figure}

For implementing an analog quantum simulation of the false vacuum
we propose to use a spinor condensate with a suppressed inter-state
scattering length $a_{12}$. As an example of this, a two-component
condensate~\cite{Egorov2011} of $^{41}$K atoms prepared in two
Zeeman states $|1\rangle=|F=1,m_{F}=1\rangle$ and $|2\rangle=|F=1,m_{F}=0\rangle$
is predicted to have an inter-state Feshbach resonance with $a_{12}\approx0$
at $675.3\,\mathrm{G}$~\cite{Lysebo2010}. States $|1\rangle$ and
$|2\rangle$ are separated by $61.93\,\mathrm{MHz}$ at this magnetic
field and are coupled via a magnetic dipole transition. The two intra-state
scattering lengths are $a_{11}=59.5a_{0}$ and $a_{22}=60.5a_{0}$,
where $a_{0}$ is the Bohr radius. The stretched state $|1\rangle$
will be condensed in the optical dipole trap. The s-wave scattering
length $a_{11}$ has a favourable value for the fast thermalization
process and is not too large to introduce inelastic losses. A pulsed
radiofrequency field of $51.63\thinspace{\rm MHz}$ will generate
the 50:50 superposition of two states $|1\rangle$ and$|2\rangle$. 

A toroidal or flat linear atom trap with tight transverse confinement
($\sim10\,\mathrm{kHz}$) will provide the 1D system with the desired
initial uniform distribution of the atom density along the axial coordinate.
Simulation data shown in Fig.~\ref{fig:1d-bubbles} and Fig.~\ref{fig:1d-fitting}
corresponds to a 1D toroidal trap. A similar experiment is also feasible
in 2D or 3D, using experimentally realized 2D~\cite{Hadzibabic2008}
and uniform 3D~\cite{Gaunt2013} trapping potentials. The results
of the corresponding simulations with realistic experimental parameters
are shown in Figs.~\ref{fig:2d-bubbles} and~\ref{fig:3d-bubbles}.
The 2D simulations of Fig.~\ref{fig:2d-bubbles} show the nucleation
of near spherical bubbles and demonstrate both the formation of domain
walls and 2D oscillons, i.e. long-lived localized non-topological
structures. The 3D simulations reveal even more complex dynamics with
multiple nested bubbles having novel topological structure seen in
Fig.~\ref{fig:3d-bubbles}a. More detailed simulations and results
from prospective experiments may further elucidate the nature of complex
bubble structures and questions like the prevalence of asymmetry in
bubble creation. 

Demonstrating the false vacuum decay by quantum tunneling will pave
the way to analog quantum simulations of a cosmological process that
is currently not accessible to exact computer simulation. Combined
with accurate observational data of the correlations in the cosmic
microwave background, this may eventually help us to refine cosmological
models and answer the question ``Where do we come from?''

\textbf{Acknowledgments:} We are indebted to Jeremy Mould and Richard
Easther for a critical reading of the manuscript and to Grigory Volovik
for pointing out Ref.~\cite{Tye2011}. This work has been supported
by the Marsden Fund of New Zealand (contract Nos.~MAU1205 and UOO1320),
the Australian Research Council, the National Science Foundation under
Grant No. PHYS-1066293 and the hospitality of the Aspen Center for
Physics.

\bibliographystyle{apsrev4-1}
\bibliography{QT}

\begin{thebibliography}{36}%
\makeatletter
\providecommand \@ifxundefined [1]{%
 \@ifx{#1\undefined}
}%
\providecommand \@ifnum [1]{%
 \ifnum #1\expandafter \@firstoftwo
 \else \expandafter \@secondoftwo
 \fi
}%
\providecommand \@ifx [1]{%
 \ifx #1\expandafter \@firstoftwo
 \else \expandafter \@secondoftwo
 \fi
}%
\providecommand \natexlab [1]{#1}%
\providecommand \enquote  [1]{``#1''}%
\providecommand \bibnamefont  [1]{#1}%
\providecommand \bibfnamefont [1]{#1}%
\providecommand \citenamefont [1]{#1}%
\providecommand \href@noop [0]{\@secondoftwo}%
\providecommand \href [0]{\begingroup \@sanitize@url \@href}%
\providecommand \@href[1]{\@@startlink{#1}\@@href}%
\providecommand \@@href[1]{\endgroup#1\@@endlink}%
\providecommand \@sanitize@url [0]{\catcode `\\12\catcode `\$12\catcode
  `\&12\catcode `\#12\catcode `\^12\catcode `\_12\catcode `\%12\relax}%
\providecommand \@@startlink[1]{}%
\providecommand \@@endlink[0]{}%
\providecommand \url  [0]{\begingroup\@sanitize@url \@url }%
\providecommand \@url [1]{\endgroup\@href {#1}{\urlprefix }}%
\providecommand \urlprefix  [0]{URL }%
\providecommand \Eprint [0]{\href }%
\providecommand \doibase [0]{http://dx.doi.org/}%
\providecommand \selectlanguage [0]{\@gobble}%
\providecommand \bibinfo  [0]{\@secondoftwo}%
\providecommand \bibfield  [0]{\@secondoftwo}%
\providecommand \translation [1]{[#1]}%
\providecommand \BibitemOpen [0]{}%
\providecommand \bibitemStop [0]{}%
\providecommand \bibitemNoStop [0]{.\EOS\space}%
\providecommand \EOS [0]{\spacefactor3000\relax}%
\providecommand \BibitemShut  [1]{\csname bibitem#1\endcsname}%
\let\auto@bib@innerbib\@empty
\bibitem [{\citenamefont {Coleman}(1977)}]{Coleman1977}%
  \BibitemOpen
  \bibfield  {author} {\bibinfo {author} {\bibfnamefont {S.}~\bibnamefont
  {Coleman}},\ }\href {\doibase 10.1103/PhysRevD.15.2929} {\bibfield  {journal}
  {\bibinfo  {journal} {Phys. Rev. D}\ }\textbf {\bibinfo {volume} {15}},\
  \bibinfo {pages} {2929} (\bibinfo {year} {1977})}\BibitemShut {NoStop}%
\bibitem [{\citenamefont {Langer}(1967)}]{Langer1967a}%
  \BibitemOpen
  \bibfield  {author} {\bibinfo {author} {\bibfnamefont {J.}~\bibnamefont
  {Langer}},\ }\href {\doibase 10.1016/0003-4916(67)90200-X} {\bibfield
  {journal} {\bibinfo  {journal} {Ann. Phys. (N. Y).}\ }\textbf {\bibinfo
  {volume} {41}},\ \bibinfo {pages} {108} (\bibinfo {year} {1967})}\BibitemShut
  {NoStop}%
\bibitem [{\citenamefont {Schmelzer}(2005)}]{Schmelzer2005}%
  \BibitemOpen
  \bibinfo {editor} {\bibfnamefont {J.~W.~P.}\ \bibnamefont {Schmelzer}},\
  ed.,\ \href {\doibase 10.1002/3527604790} {\emph {\bibinfo {title}
  {{Nucleation Theory and Applications}}}}\ (\bibinfo  {publisher} {Wiley-VCH
  Verlag GmbH \& Co. KGaA},\ \bibinfo {address} {Weinheim, FRG},\ \bibinfo
  {year} {2005})\BibitemShut {NoStop}%
\bibitem [{\citenamefont {Vilenkin}(1983)}]{Vilenkin1983}%
  \BibitemOpen
  \bibfield  {author} {\bibinfo {author} {\bibfnamefont {A.}~\bibnamefont
  {Vilenkin}},\ }\href {\doibase 10.1103/PhysRevD.27.2848} {\bibfield
  {journal} {\bibinfo  {journal} {Phys. Rev. D}\ }\textbf {\bibinfo {volume}
  {27}},\ \bibinfo {pages} {2848} (\bibinfo {year} {1983})}\BibitemShut
  {NoStop}%
\bibitem [{\citenamefont {Guth}(2007)}]{Guth2007}%
  \BibitemOpen
  \bibfield  {author} {\bibinfo {author} {\bibfnamefont {A.~H.}\ \bibnamefont
  {Guth}},\ }\href {\doibase 10.1088/1751-8113/40/25/S25} {\bibfield  {journal}
  {\bibinfo  {journal} {J. Phys. A: Math. Theor.}\ }\textbf {\bibinfo {volume}
  {40}},\ \bibinfo {pages} {6811} (\bibinfo {year} {2007})}\BibitemShut
  {NoStop}%
\bibitem [{\citenamefont {Coleman}\ and\ \citenamefont {{De
  Luccia}}(1980)}]{Coleman1980}%
  \BibitemOpen
  \bibfield  {author} {\bibinfo {author} {\bibfnamefont {S.}~\bibnamefont
  {Coleman}}\ and\ \bibinfo {author} {\bibfnamefont {F.}~\bibnamefont {{De
  Luccia}}},\ }\href {\doibase 10.1103/PhysRevD.21.3305} {\bibfield  {journal}
  {\bibinfo  {journal} {Phys. Rev. D}\ }\textbf {\bibinfo {volume} {21}},\
  \bibinfo {pages} {3305} (\bibinfo {year} {1980})}\BibitemShut {NoStop}%
\bibitem [{\citenamefont {Feeney}\ \emph {et~al.}(2011)\citenamefont {Feeney},
  \citenamefont {Johnson}, \citenamefont {Mortlock},\ and\ \citenamefont
  {Peiris}}]{Feeney2011-inflation}%
  \BibitemOpen
  \bibfield  {author} {\bibinfo {author} {\bibfnamefont {S.~M.}\ \bibnamefont
  {Feeney}}, \bibinfo {author} {\bibfnamefont {M.~C.}\ \bibnamefont {Johnson}},
  \bibinfo {author} {\bibfnamefont {D.~J.}\ \bibnamefont {Mortlock}}, \ and\
  \bibinfo {author} {\bibfnamefont {H.~V.}\ \bibnamefont {Peiris}},\ }\href
  {\doibase 10.1103/PhysRevLett.107.071301} {\bibfield  {journal} {\bibinfo
  {journal} {Phys. Rev. Lett.}\ }\textbf {\bibinfo {volume} {107}},\ \bibinfo
  {pages} {071301} (\bibinfo {year} {2011})}\BibitemShut {NoStop}%
\bibitem [{\citenamefont {Bousso}\ \emph {et~al.}(2013)\citenamefont {Bousso},
  \citenamefont {Harlow},\ and\ \citenamefont {Senatore}}]{Bousso2013}%
  \BibitemOpen
  \bibfield  {author} {\bibinfo {author} {\bibfnamefont {R.}~\bibnamefont
  {Bousso}}, \bibinfo {author} {\bibfnamefont {D.}~\bibnamefont {Harlow}}, \
  and\ \bibinfo {author} {\bibfnamefont {L.}~\bibnamefont {Senatore}},\ }\href
  {http://arxiv.org/abs/1309.4060} {\enquote {\bibinfo {title} {{Inflation
  after False Vacuum Decay: Observational Prospects after Planck}},}\ }\bibinfo
  {howpublished} {arXiv:1309.4060} (\bibinfo {year} {2013})\BibitemShut
  {NoStop}%
\bibitem [{\citenamefont {Callan}\ and\ \citenamefont
  {Coleman}(1977)}]{Callan1977}%
  \BibitemOpen
  \bibfield  {author} {\bibinfo {author} {\bibfnamefont {C.~G.}\ \bibnamefont
  {Callan}}\ and\ \bibinfo {author} {\bibfnamefont {S.}~\bibnamefont
  {Coleman}},\ }\href {\doibase 10.1103/PhysRevD.16.1762} {\bibfield  {journal}
  {\bibinfo  {journal} {Phys. Rev. D}\ }\textbf {\bibinfo {volume} {16}},\
  \bibinfo {pages} {1762} (\bibinfo {year} {1977})}\BibitemShut {NoStop}%
\bibitem [{\citenamefont {Drummond}\ and\ \citenamefont
  {Kinsler}(1989)}]{Drummond1989}%
  \BibitemOpen
  \bibfield  {author} {\bibinfo {author} {\bibfnamefont {P.~D.}\ \bibnamefont
  {Drummond}}\ and\ \bibinfo {author} {\bibfnamefont {P.}~\bibnamefont
  {Kinsler}},\ }\href {\doibase 10.1103/PhysRevA.40.4813} {\bibfield  {journal}
  {\bibinfo  {journal} {Phys. Rev. A}\ }\textbf {\bibinfo {volume} {40}},\
  \bibinfo {pages} {4813} (\bibinfo {year} {1989})}\BibitemShut {NoStop}%
\bibitem [{\citenamefont {Georgescu}\ \emph {et~al.}(2014)\citenamefont
  {Georgescu}, \citenamefont {Ashhab},\ and\ \citenamefont
  {Nori}}]{Georgescu2014}%
  \BibitemOpen
  \bibfield  {author} {\bibinfo {author} {\bibfnamefont {I.~M.}\ \bibnamefont
  {Georgescu}}, \bibinfo {author} {\bibfnamefont {S.}~\bibnamefont {Ashhab}}, \
  and\ \bibinfo {author} {\bibfnamefont {F.}~\bibnamefont {Nori}},\ }\href@noop
  {} {\bibfield  {journal} {\bibinfo  {journal} {Rev. Mod. Phys.}\ }\textbf
  {\bibinfo {volume} {86}},\ \bibinfo {pages} {153} (\bibinfo {year}
  {2014})}\BibitemShut {NoStop}%
\bibitem [{\citenamefont {Egorov}\ \emph {et~al.}(2011)\citenamefont {Egorov}
  \emph {et~al.}}]{Egorov2011}%
  \BibitemOpen
  \bibfield  {author} {\bibinfo {author} {\bibfnamefont {M.}~\bibnamefont
  {Egorov}} \emph {et~al.},\ }\href {\doibase 10.1103/PhysRevA.84.021605}
  {\bibfield  {journal} {\bibinfo  {journal} {Phys. Rev. A}\ }\textbf {\bibinfo
  {volume} {84}},\ \bibinfo {pages} {021605(R)} (\bibinfo {year}
  {2011})}\BibitemShut {NoStop}%
\bibitem [{\citenamefont {Opanchuk}\ \emph {et~al.}(2013)\citenamefont
  {Opanchuk}, \citenamefont {Polkinghorne}, \citenamefont {Fialko},
  \citenamefont {Brand},\ and\ \citenamefont
  {Drummond}}]{Opanchuk2013-early-universe}%
  \BibitemOpen
  \bibfield  {author} {\bibinfo {author} {\bibfnamefont {B.}~\bibnamefont
  {Opanchuk}}, \bibinfo {author} {\bibfnamefont {R.}~\bibnamefont
  {Polkinghorne}}, \bibinfo {author} {\bibfnamefont {O.}~\bibnamefont
  {Fialko}}, \bibinfo {author} {\bibfnamefont {J.}~\bibnamefont {Brand}}, \
  and\ \bibinfo {author} {\bibfnamefont {P.~D.}\ \bibnamefont {Drummond}},\
  }\href {\doibase 10.1002/andp.201300113} {\bibfield  {journal} {\bibinfo
  {journal} {Ann. Phys.}\ }\textbf {\bibinfo {volume} {525}},\ \bibinfo {pages}
  {866} (\bibinfo {year} {2013})}\BibitemShut {NoStop}%
\bibitem [{\citenamefont {Kapitza}(1951)}]{Kapitza1951-JETP}%
  \BibitemOpen
  \bibfield  {author} {\bibinfo {author} {\bibfnamefont {P.~L.}\ \bibnamefont
  {Kapitza}},\ }\href@noop {} {\bibfield  {journal} {\bibinfo  {journal} {Sov.
  Phys. JETP}\ }\textbf {\bibinfo {volume} {21}},\ \bibinfo {pages} {588}
  (\bibinfo {year} {1951})},\ \bibinfo {note} {for previous applications of the
  idea to BECs see H.\ Saito and M.\ Ueda, Phys.\ Rev.\ Lett.\ \textbf{90},
  040403 (2003) and H.\ Saito, R.\ G.\ Hulet, and M.\ Ueda, Phys.\ Rev.\ A
  \textbf{76}, 053619 (2007)}\BibitemShut {NoStop}%
\bibitem [{\citenamefont {Kaurov}\ and\ \citenamefont
  {Kuklov}(2005)}]{Kaurov2005}%
  \BibitemOpen
  \bibfield  {author} {\bibinfo {author} {\bibfnamefont {V.~M.}\ \bibnamefont
  {Kaurov}}\ and\ \bibinfo {author} {\bibfnamefont {A.~B.}\ \bibnamefont
  {Kuklov}},\ }\href {\doibase 10.1103/PhysRevA.71.011601} {\bibfield
  {journal} {\bibinfo  {journal} {Phys. Rev. A}\ }\textbf {\bibinfo {volume}
  {71}},\ \bibinfo {pages} {011601} (\bibinfo {year} {2005})}\BibitemShut
  {NoStop}%
\bibitem [{\citenamefont {Gritsev}\ \emph {et~al.}(2007)\citenamefont
  {Gritsev}, \citenamefont {Polkovnikov},\ and\ \citenamefont
  {Demler}}]{Gritsev2007}%
  \BibitemOpen
  \bibfield  {author} {\bibinfo {author} {\bibfnamefont {V.}~\bibnamefont
  {Gritsev}}, \bibinfo {author} {\bibfnamefont {A.}~\bibnamefont
  {Polkovnikov}}, \ and\ \bibinfo {author} {\bibfnamefont {E.}~\bibnamefont
  {Demler}},\ }\href {\doibase 10.1103/PhysRevB.75.174511} {\bibfield
  {journal} {\bibinfo  {journal} {Phys. Rev. B}\ }\textbf {\bibinfo {volume}
  {75}},\ \bibinfo {pages} {174511} (\bibinfo {year} {2007})}\BibitemShut
  {NoStop}%
\bibitem [{\citenamefont {Brand}\ \emph {et~al.}(2009)\citenamefont {Brand},
  \citenamefont {Haigh},\ and\ \citenamefont {Z\"{u}licke}}]{Brand2009}%
  \BibitemOpen
  \bibfield  {author} {\bibinfo {author} {\bibfnamefont {J.}~\bibnamefont
  {Brand}}, \bibinfo {author} {\bibfnamefont {T.~J.}\ \bibnamefont {Haigh}}, \
  and\ \bibinfo {author} {\bibfnamefont {U.}~\bibnamefont {Z\"{u}licke}},\
  }\href {\doibase 10.1103/PhysRevA.80.011602} {\bibfield  {journal} {\bibinfo
  {journal} {Phys. Rev. A}\ }\textbf {\bibinfo {volume} {80}},\ \bibinfo
  {pages} {011602} (\bibinfo {year} {2009})}\BibitemShut {NoStop}%
\bibitem [{\citenamefont {Fischer}\ and\ \citenamefont
  {Sch\"{u}tzhold}(2004)}]{Fischer2004}%
  \BibitemOpen
  \bibfield  {author} {\bibinfo {author} {\bibfnamefont {U.~R.}\ \bibnamefont
  {Fischer}}\ and\ \bibinfo {author} {\bibfnamefont {R.}~\bibnamefont
  {Sch\"{u}tzhold}},\ }\href {\doibase 10.1103/PhysRevA.70.063615} {\bibfield
  {journal} {\bibinfo  {journal} {Phys. Rev. A}\ }\textbf {\bibinfo {volume}
  {70}},\ \bibinfo {pages} {063615} (\bibinfo {year} {2004})}\BibitemShut
  {NoStop}%
\bibitem [{\citenamefont {Menicucci}\ \emph {et~al.}(2010)\citenamefont
  {Menicucci}, \citenamefont {Olson},\ and\ \citenamefont
  {Milburn}}]{Menicucci2010}%
  \BibitemOpen
  \bibfield  {author} {\bibinfo {author} {\bibfnamefont {N.~C.}\ \bibnamefont
  {Menicucci}}, \bibinfo {author} {\bibfnamefont {S.~J.}\ \bibnamefont
  {Olson}}, \ and\ \bibinfo {author} {\bibfnamefont {G.~J.}\ \bibnamefont
  {Milburn}},\ }\href {\doibase 10.1088/1367-2630/12/9/095019} {\bibfield
  {journal} {\bibinfo  {journal} {New J. Phys.}\ }\textbf {\bibinfo {volume}
  {12}},\ \bibinfo {pages} {095019} (\bibinfo {year} {2010})}\BibitemShut
  {NoStop}%
\bibitem [{\citenamefont {Neuenhahn}\ \emph {et~al.}(2012)\citenamefont
  {Neuenhahn}, \citenamefont {Polkovnikov},\ and\ \citenamefont
  {Marquardt}}]{Neuenhahn2012-phase-structures}%
  \BibitemOpen
  \bibfield  {author} {\bibinfo {author} {\bibfnamefont {C.}~\bibnamefont
  {Neuenhahn}}, \bibinfo {author} {\bibfnamefont {A.}~\bibnamefont
  {Polkovnikov}}, \ and\ \bibinfo {author} {\bibfnamefont {F.}~\bibnamefont
  {Marquardt}},\ }\href {\doibase 10.1103/PhysRevLett.109.085304} {\bibfield
  {journal} {\bibinfo  {journal} {Phys. Rev. Lett.}\ }\textbf {\bibinfo
  {volume} {109}},\ \bibinfo {pages} {085304} (\bibinfo {year}
  {2012})}\BibitemShut {NoStop}%
\bibitem [{\citenamefont {Amin}\ \emph {et~al.}(2012)\citenamefont {Amin},
  \citenamefont {Easther}, \citenamefont {Finkel}, \citenamefont {Flauger},\
  and\ \citenamefont {Hertzberg}}]{Amin2012}%
  \BibitemOpen
  \bibfield  {author} {\bibinfo {author} {\bibfnamefont {M.~A.}\ \bibnamefont
  {Amin}}, \bibinfo {author} {\bibfnamefont {R.}~\bibnamefont {Easther}},
  \bibinfo {author} {\bibfnamefont {H.}~\bibnamefont {Finkel}}, \bibinfo
  {author} {\bibfnamefont {R.}~\bibnamefont {Flauger}}, \ and\ \bibinfo
  {author} {\bibfnamefont {M.~P.}\ \bibnamefont {Hertzberg}},\ }\href {\doibase
  10.1103/PhysRevLett.108.241302} {\bibfield  {journal} {\bibinfo  {journal}
  {Phys. Rev. Lett.}\ }\textbf {\bibinfo {volume} {108}},\ \bibinfo {pages}
  {241302} (\bibinfo {year} {2012})}\BibitemShut {NoStop}%
\bibitem [{\citenamefont {B\"{a}uerle}\ \emph {et~al.}(1996)\citenamefont
  {B\"{a}uerle}, \citenamefont {Bunkov}, \citenamefont {Fisher}, \citenamefont
  {Godfrin},\ and\ \citenamefont {Pickett}}]{Bauerle1996}%
  \BibitemOpen
  \bibfield  {author} {\bibinfo {author} {\bibfnamefont {C.}~\bibnamefont
  {B\"{a}uerle}}, \bibinfo {author} {\bibfnamefont {Y.~M.}\ \bibnamefont
  {Bunkov}}, \bibinfo {author} {\bibfnamefont {S.~N.}\ \bibnamefont {Fisher}},
  \bibinfo {author} {\bibfnamefont {H.}~\bibnamefont {Godfrin}}, \ and\
  \bibinfo {author} {\bibfnamefont {G.~R.}\ \bibnamefont {Pickett}},\ }\href
  {\doibase 10.1038/382332a0} {\bibfield  {journal} {\bibinfo  {journal}
  {Nature}\ }\textbf {\bibinfo {volume} {382}},\ \bibinfo {pages} {332}
  (\bibinfo {year} {1996})}\BibitemShut {NoStop}%
\bibitem [{\citenamefont {Volovik}(2009)}]{Volovik2009}%
  \BibitemOpen
  \bibfield  {author} {\bibinfo {author} {\bibfnamefont {G.~E.}\ \bibnamefont
  {Volovik}},\ }\href@noop {} {\emph {\bibinfo {title} {{The Universe in a
  Helium Droplet}}}}\ (\bibinfo  {publisher} {OUP Oxford},\ \bibinfo {year}
  {2009})\ p.\ \bibinfo {pages} {511}\BibitemShut {NoStop}%
\bibitem [{\citenamefont {Satoh}\ \emph {et~al.}(1992)\citenamefont {Satoh},
  \citenamefont {Morishita}, \citenamefont {Ogata},\ and\ \citenamefont
  {Katoh}}]{Satoh1992}%
  \BibitemOpen
  \bibfield  {author} {\bibinfo {author} {\bibfnamefont {T.}~\bibnamefont
  {Satoh}}, \bibinfo {author} {\bibfnamefont {M.}~\bibnamefont {Morishita}},
  \bibinfo {author} {\bibfnamefont {M.}~\bibnamefont {Ogata}}, \ and\ \bibinfo
  {author} {\bibfnamefont {S.}~\bibnamefont {Katoh}},\ }\href {\doibase
  10.1103/PhysRevLett.69.335} {\bibfield  {journal} {\bibinfo  {journal} {Phys.
  Rev. Lett.}\ }\textbf {\bibinfo {volume} {69}},\ \bibinfo {pages} {335}
  (\bibinfo {year} {1992})}\BibitemShut {NoStop}%
\bibitem [{\citenamefont {Tye}\ and\ \citenamefont {Wohns}(2011)}]{Tye2011}%
  \BibitemOpen
  \bibfield  {author} {\bibinfo {author} {\bibfnamefont {S.-H.~H.}\
  \bibnamefont {Tye}}\ and\ \bibinfo {author} {\bibfnamefont {D.}~\bibnamefont
  {Wohns}},\ }\href {\doibase 10.1103/PhysRevB.84.184518} {\bibfield  {journal}
  {\bibinfo  {journal} {Phys. Rev. B}\ }\textbf {\bibinfo {volume} {84}},\
  \bibinfo {pages} {184518} (\bibinfo {year} {2011})}\BibitemShut {NoStop}%
\bibitem [{\citenamefont {Steel}\ \emph {et~al.}(1998)\citenamefont {Steel}
  \emph {et~al.}}]{Steel1998}%
  \BibitemOpen
  \bibfield  {author} {\bibinfo {author} {\bibfnamefont {M.}~\bibnamefont
  {Steel}} \emph {et~al.},\ }\href {\doibase 10.1103/PhysRevA.58.4824}
  {\bibfield  {journal} {\bibinfo  {journal} {Phys. Rev. A}\ }\textbf {\bibinfo
  {volume} {58}},\ \bibinfo {pages} {4824} (\bibinfo {year}
  {1998})}\BibitemShut {NoStop}%
\bibitem [{\citenamefont {Sinatra}\ \emph {et~al.}(2002)\citenamefont
  {Sinatra}, \citenamefont {Lobo},\ and\ \citenamefont {Castin}}]{Sinatra2002}%
  \BibitemOpen
  \bibfield  {author} {\bibinfo {author} {\bibfnamefont {A.}~\bibnamefont
  {Sinatra}}, \bibinfo {author} {\bibfnamefont {C.}~\bibnamefont {Lobo}}, \
  and\ \bibinfo {author} {\bibfnamefont {Y.}~\bibnamefont {Castin}},\ }\href
  {\doibase 10.1088/0953-4075/35/17/301} {\bibfield  {journal} {\bibinfo
  {journal} {J. Phys. B}\ }\textbf {\bibinfo {volume} {35}},\ \bibinfo {pages}
  {3599} (\bibinfo {year} {2002})}\BibitemShut {NoStop}%
\bibitem [{\citenamefont {Caldeira}\ and\ \citenamefont
  {Leggett}(1981)}]{Caldeira1981}%
  \BibitemOpen
  \bibfield  {author} {\bibinfo {author} {\bibfnamefont {A.~O.}\ \bibnamefont
  {Caldeira}}\ and\ \bibinfo {author} {\bibfnamefont {A.~J.}\ \bibnamefont
  {Leggett}},\ }\href {\doibase 10.1103/PhysRevLett.46.211} {\bibfield
  {journal} {\bibinfo  {journal} {Phys. Rev. Lett.}\ }\textbf {\bibinfo
  {volume} {46}},\ \bibinfo {pages} {211} (\bibinfo {year} {1981})}\BibitemShut
  {NoStop}%
\bibitem [{\citenamefont {Atland}\ and\ \citenamefont
  {Simons}(2010)}]{Atland2010}%
  \BibitemOpen
  \bibfield  {author} {\bibinfo {author} {\bibfnamefont {A.}~\bibnamefont
  {Atland}}\ and\ \bibinfo {author} {\bibfnamefont {B.}~\bibnamefont
  {Simons}},\ }\href@noop {} {\emph {\bibinfo {title} {{Condensed Matter Field
  Theory}}}}\ (\bibinfo  {publisher} {Cambridge University Press},\ \bibinfo
  {year} {2010})\ p.\ \bibinfo {pages} {783}\BibitemShut {NoStop}%
\bibitem [{\citenamefont {Liddle}\ and\ \citenamefont
  {Lyth}(2000)}]{Liddle2000}%
  \BibitemOpen
  \bibfield  {author} {\bibinfo {author} {\bibfnamefont {A.~R.}\ \bibnamefont
  {Liddle}}\ and\ \bibinfo {author} {\bibfnamefont {D.~H.}\ \bibnamefont
  {Lyth}},\ }\href
  {http://www.cambridge.org/au/academic/subjects/physics/cosmology-relativity-and-gravitation/cosmological-inflation-and-large-scale-structure?format=PB}
  {\emph {\bibinfo {title} {{Cosmological Inflation and Large-Scale
  Structure}}}}\ (\bibinfo  {publisher} {Cambridge University Press},\ \bibinfo
  {year} {2000})\ p.\ \bibinfo {pages} {400}\BibitemShut {NoStop}%
\bibitem [{\citenamefont {Vachaspati}(2006)}]{Vachaspati2006}%
  \BibitemOpen
  \bibfield  {author} {\bibinfo {author} {\bibfnamefont {T.}~\bibnamefont
  {Vachaspati}},\ }\href@noop {} {\emph {\bibinfo {title} {{Kinks and Domain
  Walls: An Introduction to Classical and Quantum Solitons}}}}\ (\bibinfo
  {publisher} {Cambridge University Press},\ \bibinfo {year}
  {2006})\BibitemShut {NoStop}%
\bibitem [{\citenamefont {Su}\ \emph {et~al.}(2014)\citenamefont {Su},
  \citenamefont {Gou}, \citenamefont {Liu}, \citenamefont {Bradley},
  \citenamefont {Fialko},\ and\ \citenamefont {Brand}}]{Su2014a}%
  \BibitemOpen
  \bibfield  {author} {\bibinfo {author} {\bibfnamefont {S.-W.}\ \bibnamefont
  {Su}}, \bibinfo {author} {\bibfnamefont {S.-C.}\ \bibnamefont {Gou}},
  \bibinfo {author} {\bibfnamefont {I.-K.}\ \bibnamefont {Liu}}, \bibinfo
  {author} {\bibfnamefont {A.~S.}\ \bibnamefont {Bradley}}, \bibinfo {author}
  {\bibfnamefont {O.}~\bibnamefont {Fialko}}, \ and\ \bibinfo {author}
  {\bibfnamefont {J.}~\bibnamefont {Brand}},\ }\href@noop {} {\enquote
  {\bibinfo {title} {Oscillons in coupled {B}ose-{E}instein condensates},}\ }
  (\bibinfo {year} {2014}),\ \bibinfo {note} {in preparation}\BibitemShut
  {NoStop}%
\bibitem [{\citenamefont {Takagi}(2006)}]{Takagi2006}%
  \BibitemOpen
  \bibfield  {author} {\bibinfo {author} {\bibfnamefont {S.}~\bibnamefont
  {Takagi}},\ }\href@noop {} {\emph {\bibinfo {title} {{Macroscopic Quantum
  Tunneling}}}}\ (\bibinfo  {publisher} {Cambridge University Press},\ \bibinfo
  {year} {2006})\ p.\ \bibinfo {pages} {224}\BibitemShut {NoStop}%
\bibitem [{\citenamefont {Gaunt}\ \emph {et~al.}(2013)\citenamefont {Gaunt},
  \citenamefont {Schmidutz}, \citenamefont {Gotlibovych}, \citenamefont
  {Smith},\ and\ \citenamefont {Hadzibabic}}]{Gaunt2013}%
  \BibitemOpen
  \bibfield  {author} {\bibinfo {author} {\bibfnamefont {A.~L.}\ \bibnamefont
  {Gaunt}}, \bibinfo {author} {\bibfnamefont {T.~F.}\ \bibnamefont
  {Schmidutz}}, \bibinfo {author} {\bibfnamefont {I.}~\bibnamefont
  {Gotlibovych}}, \bibinfo {author} {\bibfnamefont {R.~P.}\ \bibnamefont
  {Smith}}, \ and\ \bibinfo {author} {\bibfnamefont {Z.}~\bibnamefont
  {Hadzibabic}},\ }\href {\doibase 10.1103/PhysRevLett.110.200406} {\bibfield
  {journal} {\bibinfo  {journal} {Phys. Rev. Lett.}\ }\textbf {\bibinfo
  {volume} {110}},\ \bibinfo {pages} {200406} (\bibinfo {year}
  {2013})}\BibitemShut {NoStop}%
\bibitem [{\citenamefont {Lysebo}\ and\ \citenamefont
  {Veseth}(2010)}]{Lysebo2010}%
  \BibitemOpen
  \bibfield  {author} {\bibinfo {author} {\bibfnamefont {M.}~\bibnamefont
  {Lysebo}}\ and\ \bibinfo {author} {\bibfnamefont {L.}~\bibnamefont
  {Veseth}},\ }\href {\doibase 10.1103/PhysRevA.81.032702} {\bibfield
  {journal} {\bibinfo  {journal} {Phys. Rev. A}\ }\textbf {\bibinfo {volume}
  {81}},\ \bibinfo {pages} {032702} (\bibinfo {year} {2010})}\BibitemShut
  {NoStop}%
\bibitem [{\citenamefont {Hadzibabic}\ \emph {et~al.}(2008)\citenamefont
  {Hadzibabic}, \citenamefont {Kr\"{u}ger}, \citenamefont {Cheneau},
  \citenamefont {Rath},\ and\ \citenamefont {Dalibard}}]{Hadzibabic2008}%
  \BibitemOpen
  \bibfield  {author} {\bibinfo {author} {\bibfnamefont {Z.}~\bibnamefont
  {Hadzibabic}}, \bibinfo {author} {\bibfnamefont {P.}~\bibnamefont
  {Kr\"{u}ger}}, \bibinfo {author} {\bibfnamefont {M.}~\bibnamefont {Cheneau}},
  \bibinfo {author} {\bibfnamefont {S.~P.}\ \bibnamefont {Rath}}, \ and\
  \bibinfo {author} {\bibfnamefont {J.}~\bibnamefont {Dalibard}},\ }\href
  {\doibase 10.1088/1367-2630/10/4/045006} {\bibfield  {journal} {\bibinfo
  {journal} {New J. Phys.}\ }\textbf {\bibinfo {volume} {10}},\ \bibinfo
  {pages} {045006} (\bibinfo {year} {2008})}\BibitemShut {NoStop}%
\end{thebibliography}%

\clearpage{}
\end{document}